\documentclass[doublecolumn]{aa}

\usepackage{graphicx}
\usepackage{multirow}
\usepackage{txfonts}
\begin{document} 


   \title{Extreme Fluctuations in the Sun's Activity over the Modern Maximum: Understanding the Enigmatic Solar Cycles 19-20}

   \author{S. Pal
          \inst{1}
          \and
          D. Nandy\inst{1,2}
          }

   \institute{Center of Excellence in Space Sciences India, Indian Institute of Science Education and Research Kolkata, Mohanpur 741246, West Bengal, India\\
              \email{shao.physics@gmail.com}
         \and
             Department of Physical Sciences, Indian Institute of Science Education and Research Kolkata, Mohanpur 741246, West Bengal, India\\
             \email{dnandi@iiserkol.ac.in}\\
             }

  \abstract
   {Over the past century, the Sun's activity -- which exhibits significant variations -- went through a phase known as the Modern Maximum. Notably, the strongest sunspot cycle on record during this period, and indeed since direct sunspot observations began, was cycle 19; this was followed by a significantly weaker cycle 20. Understanding and reconstructing this extreme variability has remained elusive. Utilizing data-driven, coupled models of magnetic field evolution on the Sun's surface and within its convection zone, here we show that random deviations in the tilt angle and polarity orientation of bipolar sunspot pairs is sufficient to explain these observed, extreme fluctuations during the modern maximum in solar activity. Our results support the theory that perturbation in the poloidal field source of the dynamo mechanism -- mediated via the emergence of anomalously tilted solar active regions - is the primary driver of extreme variations in the Sun's activity. This study has implications for understanding how the Sun may switch from a phase of extreme activity to quiescent, low activity phases -- such as the Maunder Minimum.}


   \titlerunning{Origin of Extreme Solar Cycle Fluctuations}
   
   \maketitle
%

\section{Introduction}

The sunspot cycle -- characterized by an approximately 11-year quasi-periodic rise and fall in solar activity -- is a striking manifestation of the Sun's magnetic behavior \citep{Charbonneau2020, Usoskin2023}. However, the strength of the Sun's activity cycle is not uniform and varies from one cycle to another, resulting in a variable forcing of the heliosphere that seamlessly bridges physical phenomena originating in our host star's interior to planetary impacts \citep{Daglis2021}. Observational evidence has uncovered that over the past century, the Sun has exhibited a prolonged period of unusually high activity known as the {\it Modern Maximum} \citep{Solanki2004}. The strongest cycle during this phase -- in fact over the past four centuries since direct sunspot cycle observations began with Galileo Galilei and his contemporaries -- was sunspot cycle 19 peaking around 1957; this was followed by an unexpectedly weak sunspot cycle 20. Understanding and reconstructing this extreme solar activity fluctuation over cycles 19-20 through physics-based models has remained an outstanding challenge \citep{Cameron2010, Jiang2013, Bhowmik2018, Virtanen2022, Pal2024, Yeates2025ApJ}. 

\begin{figure*}
\centering
\includegraphics[scale=0.35]{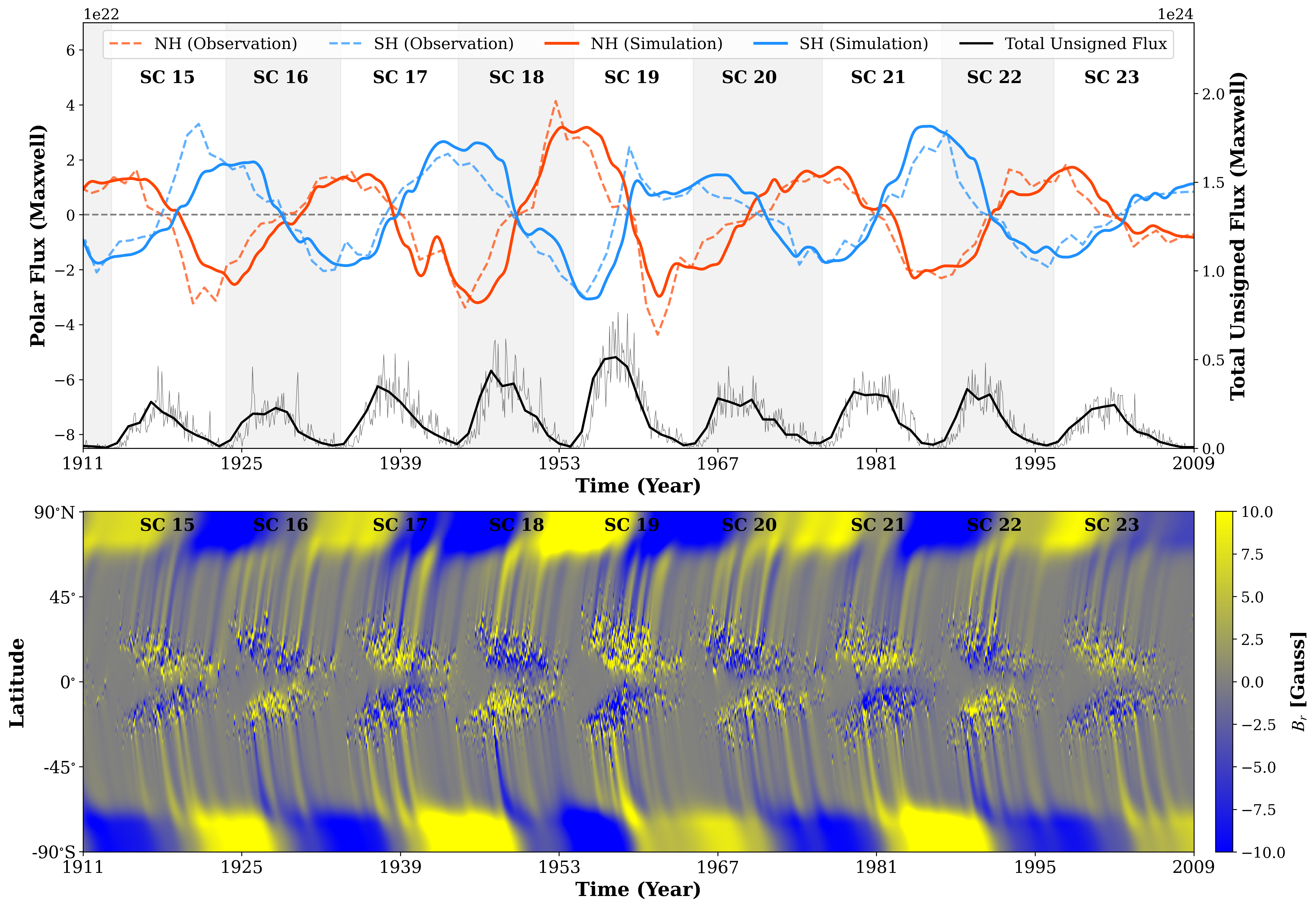}
\caption{The top panel illustrates the variations in the optimized simulated polar flux (solid curves) alongside the observed polar flux obtained from the MWO polar faculae database (dashed curves) for solar cycles 15 to 23. The red and blue curves represent the Northern and Southern hemispheres, respectively. Additionally, the black/grey curve in the same panel depicts the total yearly/monthly averaged unsigned sunspot flux, derived from the RGO/USAF/NOAA database. The bottom panel presents the time-latitudinal distribution of the radial magnetic field ($\mathrm{B_r}$) based on the optimized simulation. Here, yellow and blue shades indicate magnetic fields of positive and negative polarity, respectively.}\label{fig:1}

\end{figure*}

It is understood that nonlinearities inherent in the magnetohydrodynamic solar dynamo mechanism and stochastic perturbations in the dynamo source terms can lead to amplitude fluctuations from one solar cycle to another \citep{Charbonneau2000, Saha2025}. The Sun's large-scale dipolar field -- the poloidal component (of which the polar field is a proxy) -- is inducted by solar differential rotation within the Sun's convection zone to produce the toroidal magnetic field component of the following sunspot cycle \citep{Parker1955}. Magnetic (Lorentz) feedback of strong toroidal flux tubes on the Sun's differential rotation can, in principle, result in amplitude modulation; however, observations show that inter-cycle variations (known as torsional oscillations) in the solar differential rotation which acts as the source of the Sun's toroidal field -- is quite small $\leq$ 5$\%$ \citep{Mahajan2024SoPh}. 

Strong toroidal fields are unstable to magnetic buoyancy and rise up to emerge through the solar surface giving rise to bipolar sunspot pairs --- which are observed to be systematically tilted relative to the local latitude (due to the action of the Coriolis force on rising flux tubes). Observations \citep{Dasi2010, Munoz2012}, theoretical considerations \citep{Cameron2015} as well as data-driven, physical models of the long-term evolution of solar magnetic fields -- such as solar surface flux transport (SFT) models and dynamo models \citep{Bhowmik2018} -- have suggested that the dispersal of the magnetic flux of these tilted bipolar sunspots (mediated by plasma flows) is the primary mechanism for recreation of the Sun's large scale dipolar field; the latter mechanism is known as the Babcock-Leighton (BL) mechanism \citep{Babcock1961APJ, Leighton1969ApJ}. As magnetic flux tubes rise through the solar convection zone, they are buffeted by vigorous turbulence, generating an observed scatter in the tilt angles around the mean (Joy's law tilt) expected from Coriolis force \citep{Cheung2014}. The amplitude of this scatter is observed to be much more than the mean tilt angle, which acts as a significant source of perturbation on the poloidal field source \citep{Jiang2014, Nagy2017}. Anti-Hale active regions (ARs) -- bipolar sunspot pairs whose polarity orientation does not conform to the conventional solar cycle trend -- are an additional source of perturbation \citep{Nagy2017, Pal2023}. Other mechanisms, such as tilt quenching and latitudinal quenching, are also known to act as amplitude-limiting mechanisms for the BL poloidal field generation mechanism \citep{Jiao2021, Yeates2025ApJ}. 

Although the interplay of these nonlinearities and stochastic forcing are theorized to drive solar activity fluctuations, capturing the extreme variation between cycles 19-20 has remained elusive. Here we address the outstanding question whether solar cycles 19 and 20 can be accurately reconstructed using known physics and physical models of solar magnetic field evolution.

\begin{figure*}
\centering
\includegraphics[scale=0.6]{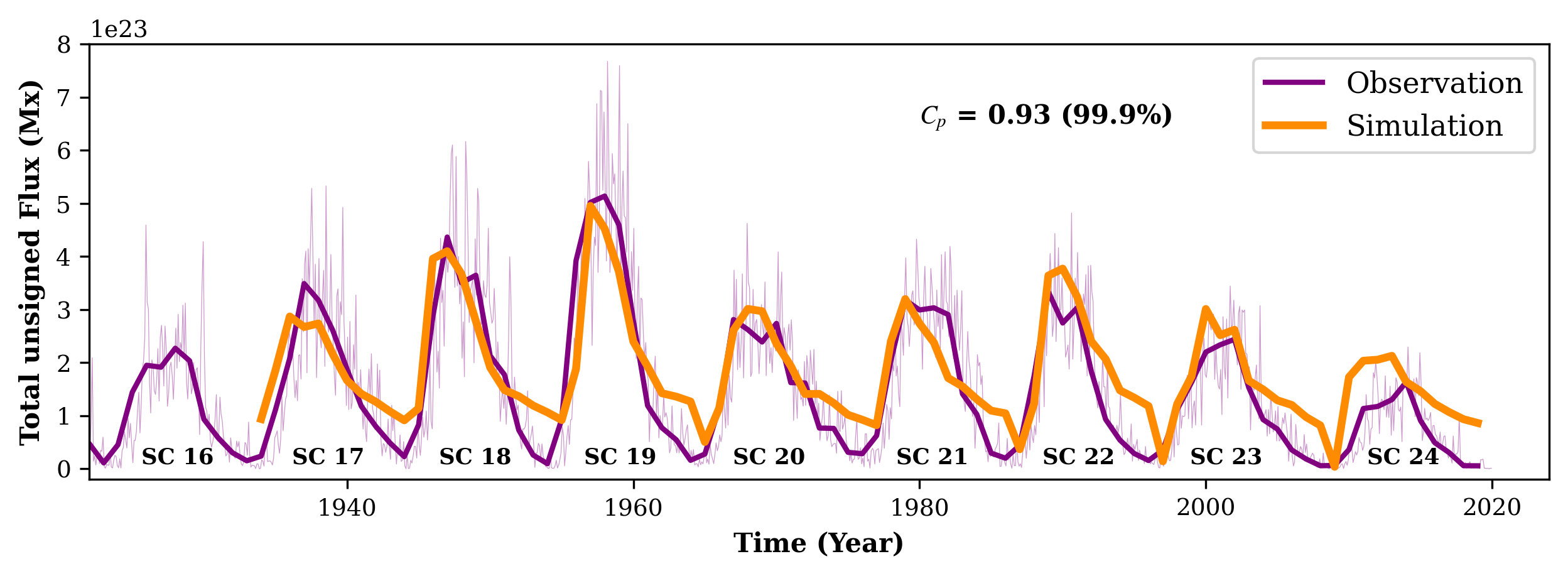}
\caption{The violet curve represents the total yearly averaged unsigned sunspot flux from solar cycle 16 to 24, derived from the USAF/RGO/NOAA database. The orange curve indicates the total unsigned flux simulated using the dynamo model (refer \ref{sec:methoddynamo}), driven by the poloidal field generated by the C-PhoTraM model. A Pearson correlation coefficient of 0.93, calculated with a 99$\%$ confidence level, highlights the strong agreement between the observed and simulated flux strengths at cycle maxima.}\label{fig:2}
\end{figure*}

\begin{figure}
\centering
\includegraphics[scale=0.55]{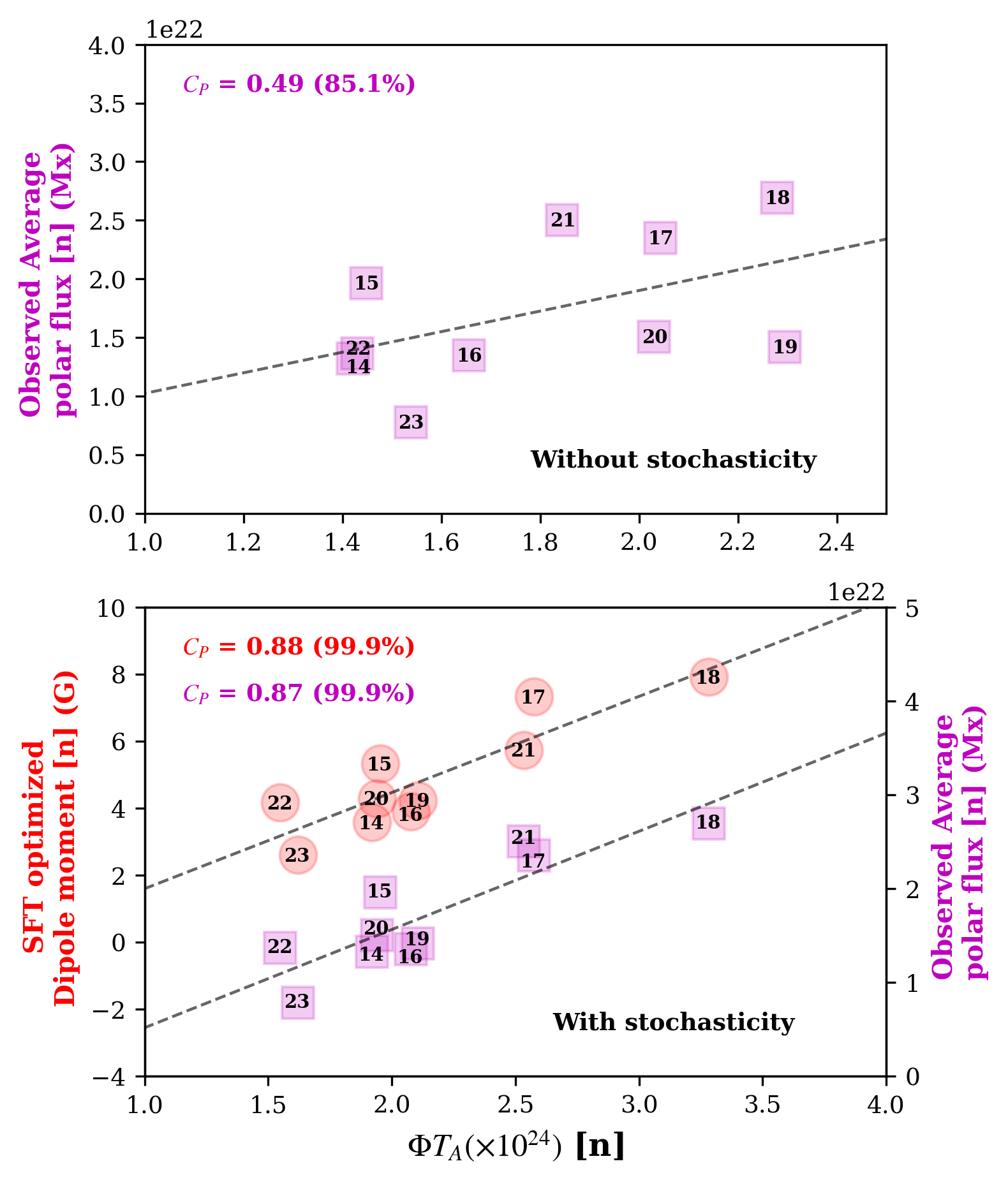}
\caption{Top panel denotes correlation between the flux-weighted tilt coefficient multiplied by total flux ($\mathrm{\Phi T_A}$) for the $\mathrm{n^{th}}$ solar cycle and the observed average polar flux at the end of the same cycle. This correlation is estimated using the sunspot database without incorporating stochasticity. Bottom panel illustrates correlation between the $\mathrm{\Phi T_A}$ and two quantities at the end of the same solar cycle ($\mathrm{n}$): (i) the dipole moment simulated by C-PhoTraM simulation (left $x$-axis, red circles), and (ii) the observed average polar flux (right $x$-axis, magenta squares). These correlations are estimated after incorporating stochasticity in the sunspot emergence statistics. The grey dashed lines in both panels represent the best-fit linear regressions.}\label{fig:3}
\end{figure}

\section{Results and Discussions} \label{sec:results1}

We first reconstruct century-scale solar polar field variations using an SFT model. These models are driven by incorporating as input the emergence profile and statistics of bipolar sunspot pairs. We prepare the synthetic input of ARs using observed sunspot emergence statistics derived from the \cite{rgo} database -- which provides details on the area, emergence time, and location of sunspots. These information are used to drive the C-PhoTraM model (a brief description of which is available in appendix section \ref{sec:methodsft}). Past work has shown that long-term simulations using this database fail to reproduce variations across all solar cycles, including extreme solar cycles \citep{Bhowmik2018, Pal2024}. This limitation arises because this database lacks two crucial information necessary for precise modeling. The first is the tilt angle information of ARs, which significantly influences polar field generation, is unavailable. Although models often assume that tilt angles follow the mean Joy’s law tilt expected from Coriolis force, scatter around the mean tilt is crucial for capturing polar field variations \citep{Jiang2014, Nagy2017}. The second limitation is the lack of observational constraints on the polarity orientation of sunspots which results in an inability to account for the impact of anomalous ARs (anti-Hale configurations). These anomalous sunspots can significantly reduce the polar field and the overall solar cycle strength \citep{Pal2023}. To overcome these limitations, we incorporate these two properties of ARs into our C-PhoTraM simulations.

We begin our simulation in 1902, at the beginning of solar cycle 14, initializing it with a dipolar magnetic field configuration. First, we generate the ensembles of sunspot input sources for a single solar cycle. Each ensemble introduces random scatter into the standard empirically calculated tilt angles; for technical details, see \ref{sec:methodsft}. Additionally, we distribute approximately $8\%$ - $8.4\%$ of the total ARs of that cycle as anti-Hale regions -- which we note is consistent with the observed range gleaned from the most recent solar cycles for which polarity information exists \citep{Li2018}.  Accounting for tilt scatter and anti-Hale active region distribution in this manner, we generate multiple input sources for solar cycles 14 to 23 and perform Monte Carlo (MC) C-PhoTraM simulations. From these ensembles, we choose the optimal solution for each solar cycle, which minimizes the difference between the simulated and observed polar field at cycle minima (for detailed methodology of optimizing, see \ref{sec:methodsft})

Figure \ref{fig:1} (top panel) shows the optimized simulated polar flux variations for the northern and southern hemispheres (solid red and blue, respectively) compared with the observed polar flux (dashed curves). This figure demonstrates that by accounting for tilt angle scatter and anti-Hale sunspots, we are able to successfully reproduce the polar field evolution of solar cycles 18 and 19 (which serve as sources for sunspot cycles 19 and 20, respectively), along with other cycles over century-scale. The butterfly diagram (Figure \ref{fig:1} bottom panel) depicts the time-latitude evolution of the longitudinally averaged radial magnetic field ($\mathrm{B_r}$). The diagram shows the surface field dynamics associated with polar field reversal and build-up in our optimized simulation. Additionally, the magnetic flux surges toward the poles, representing both positive and negative polarity contributions, including perturbation due to anomalous ARs are clearly discernible. These surges play a crucial role in determining polar field amplitude.

In the next stage, we employ a two-dimensional (axisymmetric) kinematic dynamo model (for model details see appendix section \ref{sec:methoddynamo}) to reconstruct the solar cycle time series over a century covering the Modern Maximum phase. In this approach, we assimilate the longitudinally averaged surface magnetic field from the optimized SFT simulation as the source of the poloidal field into the dynamo model at the end of each solar cycle (see appendix \ref{sec:methoddynamo} for a more detailed methodology). Our solar dynamo model integrates a buoyancy algorithm to simulate sunspot emergence as eruptions of the toroidal field when it exceeds a specified threshold magnetic field strength. These toroidal field eruptions are then used as a proxy for the total sunspot flux that has erupted during the cycle. The simulated magnetic flux from these sunspot eruptions is compared with the observed unsigned sunspot flux, which represents the total unsigned flux derived from sunspot emergence data in the RGO/USAF/NOAA database. The result is depicted in Figure \ref{fig:2}. It is noted that the simulated magnetic flux does not reach the observed low value at minima. This is a general issue in diverse dynamo models due to cycle overlap \citep{Hazra2016, Kumar2019}. This can be by fine-tuning the meridional flow from cycle to cycle, which we do not consider here \citep{Hazra2017}. Our results show that the C-PhoTraM generated poloidal field successfully reproduces the cycle strength of solar cycles 19 and 20, while capturing the overall trend of other cycles over the century scale. This outcome highlights the importance of properly accounting for tilt angle scatter and anomalous regions in governing extreme variations in solar activity.

To deconstruct the physics of extreme solar variability over sunspot cycles 19-20, we study the mean flux weighted tilt angle distribution of sunspots in the simulation runs that successfully reproduce the polar flux for solar cycles 18 and 19 (which subsequently act as seeds of sunspot cycles 19 and 20, respectively). If no stochasticity is included in tilt-angle variations, the values of lux-weighted tilt coefficient, i.e. $\mathrm{\Phi T_A}$ remain nearly identical for cycles 18 and 19, as shown in the top panel of Figure \ref{fig:3}. However, upon incorporating tilt-angle fluctuations, $\mathrm{\Phi T_A}$ increases substantially for cycle 18, while it decreases for cycle 19. Moreover, the correlation between $\mathrm{\Phi T_A}$ of the $\mathrm{n^{th}}$ solar cycle and the observed average polar flux at the end of the same cycle improves significantly when stochasticity is incorporated into the sunspot emergence statistics (compare the magenta squares in the top and bottom panels of Figure \ref{fig:3}). Similarly, we observe a strong positive correlation ($r = 0.88$, with $99.9\%$ confidence) between $\mathrm{\Phi T_A}$ of the $\mathrm{n^{th}}$ cycle and the dipole moment at the end of the same cycle, as generated by the optimized SFT simulation (see the red circles in the bottom panel of Figure \ref{fig:3}). This result suggests that solar cycle 18, with its higher $\mathrm{\Phi T_A}$ leads to a stronger dipole moment at the end of the cycle which seeds the extreme sunspot cycle 19. This is because, when the tilt angle of a sunspot pair is high, the latitudinal separation between opposite polarities increases due to plasma flows, avoiding intra-active region flux cancellation and allowing transport of significant amount of flux efficiently toward the poles. A stronger polar field enhances the seed poloidal field for the following cycle, which explains the exceptionally strong solar cycle 19. The highest value of $\mathrm{\Phi T_A}$ further indicates that cycle 18 had ARs with very high tilt-angle scatter -- or, equivalently, the highest degree of stochasticity -- accounting for its exceptionally high amplitude.

On the other hand, the sudden drop in $\mathrm{\Phi T_A}$ observed for solar cycle 19 offers a contrasting and compelling explanation for the weak amplitude of solar cycle 20. When the tilt angle of a sunspot pair is small, a larger portion of magnetic flux cancels internally within the active region itself, thereby reducing the efficiency of poleward flux transport. This diminished transport of magnetic flux weakens the buildup of the polar field, which in turn contributes to a weaker subsequent solar cycle -- as observed in cycle 20. Additionally, the $\mathrm{\Phi T_A}$ parameter accounts for the influence of anomalous sunspots, including anti-Hale ARs, which possess reversed polarities relative to standard sunspots. Being a stronger cycle, solar cycle 19 exhibits a higher number of such anomalous ARs compared to cycle 18, as shown in the Figure \ref{fig:a1} in the appendix section \ref{sec:methodsft}. These anomalous ARs contribute a greater amount of opposite-polarity flux toward the poles, ultimately diminishing the net polar field strength by the end of cycle 19. Consequently, anti-Hale ARs play a crucial role in generating extreme solar cycle variability—where a high-amplitude cycle culminates in a weak polar field, leading to a subdued following cycle.

We note that it is challenging to separately assess the effects of anti-Joy and anti-Hale regions across solar cycles 18 and 19 due to several factors which influence their contributions. The most important factor, elaborated in detail in \cite{Pal2023}, is that an anti-Joy region with a tilt angle opposite to that expected from Joy’s Law evolves into an anti-Hale region (due to the action of differential rotation) within a rotational timescale -- far shorter than solar cycle timescales -- and thus their contributions are nearly similar. Therefore, from the physical perspective, it is a better strategy to group these types of ARs within the general framework of anomalous ARs.

\section{Summary} \label{sec:summary}

Solar cycle 19 stands out as exceptionally strong, yet the polar field build-up at its end is surprisingly weak, leading to a weaker subsequent solar cycle 20. Such variations in solar cycles can arise from multiple factors, including the non-linear mechanisms and external stochastic mechanisms involved in magnetic field generation and transport. The lack of long-term observations makes it challenging to constrain these processes. In this study, we employ a novel ensemble run methodology applied on a coupled, data-driven solar surface flux transport model and dynamo model to reconstruct the past ten solar cycles. We specifically focus on understanding the extreme variation from sunspot cycle 19 to 20. 

Our results reinforce the hypothesis that one of the key factors -- if not the most important one -- driving cycle-to-cycle variations is the random scattering in sunspot tilt angles. More importantly, our simulations show that reasonable fluctuations in the tilt angle of ARs -- within the observed range of variabilities -- is able to recover the significant variation in polar field at the end of cycles 18 and 19 that act as seeds for historically strong sunspot cycle 19 and the much weaker cycle 20. Our results imply that no exotic new physics need to be invoked to explain the extreme fluctuations observed during the Modern maximum in solar activity. Our data-driven, observationally constrained physics-based simulations lend further credence to the emerging understanding that stochastic perturbations -- and not non-linear quenching -- is the primary driver of centennial-scale solar variability.

Building upon our findings, we may surmise that such random (stochastic) fluctuations, manifest in solar ARs with highly anomalous tilt (and sometimes with large flux) may indeed result in a catastrophic reduction in the value of the polar field -- precipitating a Maunder-like grand minimum. This possibility, already alluded to in \cite{Nagy2017}, appears to be a distinct possibility based on this work which recovers the extreme fall in amplitude from sunspot cycle 19 to 20. While \cite{Nagy2017} suggest a ``rogue'' or extreme anomalous AR may achieve this, a number of anomalous ARs -- realizing which is more probable -- may achieve an analogous effect in precipitating grand minima episodes. Our work motivates further investigations into intriguing possibilities. 
 
\begin{acknowledgements}
This research was conducted at the Center of Excellence in Space Sciences India (CESSI), supported by IISER Kolkata and the Ministry of Education, Government of India. We acknowledge useful discussions and feedback from Prantika Bhowmik, Paul Charbonneau and Kristóf Petrovay. S.P. thanks the Department of Technical Education, Training and Skill Development, Government of West Bengal.
\end{acknowledgements}

%
\bibliographystyle{aa} 
\bibliography{biblio} 
%


\begin{appendix} 
\section{Numerical Model Description}

\subsection{Data-driven C-PhoTraM Simulation}\label{sec:methodsft}
\begin{figure}
\centering
\includegraphics[scale=0.37]{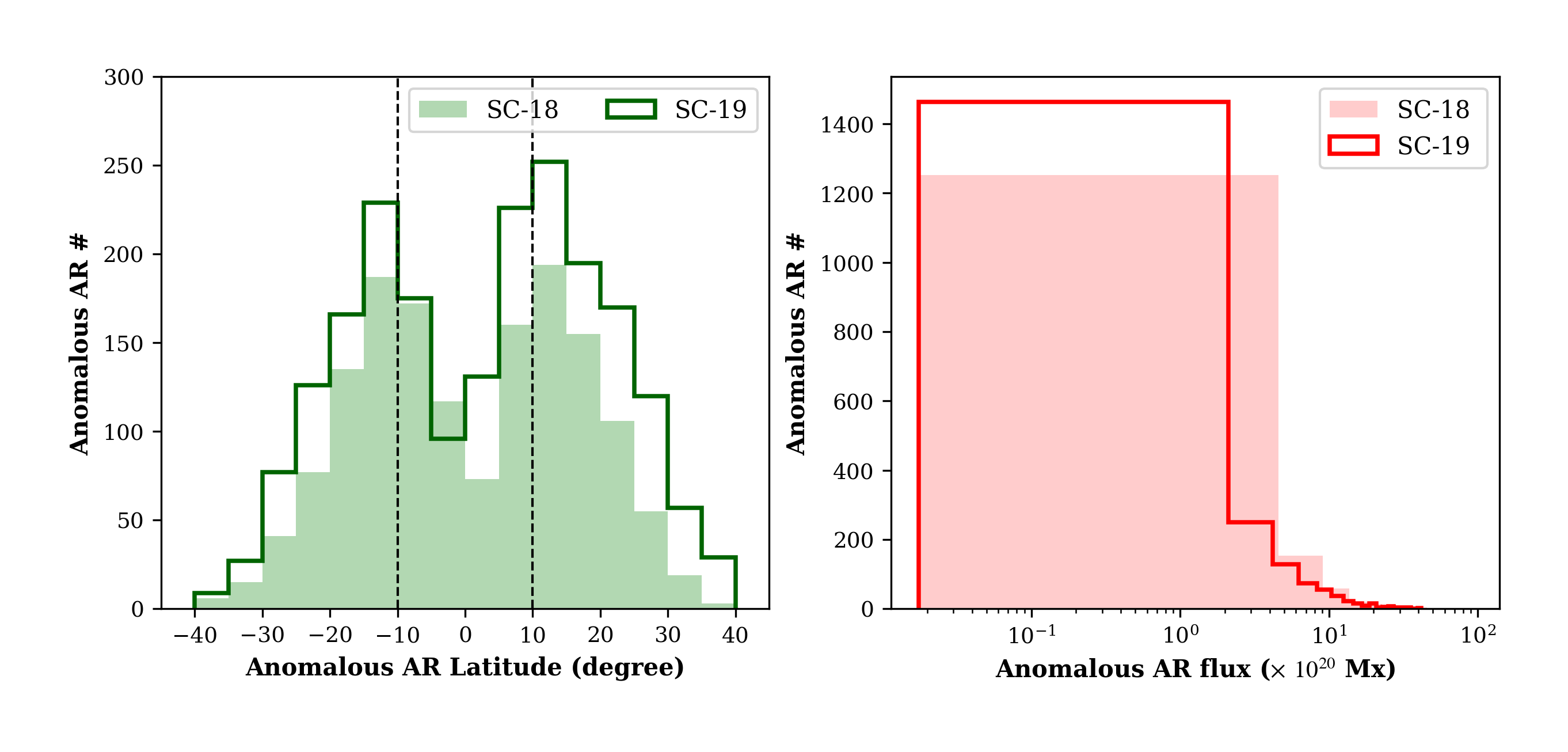}
\caption{The top panel illustrates the latitudinal distribution of anomalous ARs (both anti-Hale and anti-Joy regions) for solar cycle 18 (filled histogram) and solar cycle 19 (solid histogram). The bottom panel displays the flux distribution of anomalous sunspots, with red filled and solid histograms representing the flux throughout the sunspot cycles 18 and 19, respectively. For clarity and better visualization, the x-axis in this panel is on a logarithmic scale.}\label{fig:a1}
\end{figure}

\textbf{Model Equation:}\\
The CESSI Photospheric-Flux Transport Model (C-PhoTraM) is a newly developed two-dimensional numerical model designed to simulate the solar surface magnetic flux transport processes. It solves the radial component of the magnetic induction equation in the spatial domain in the presence of diffusion $\eta$ (mimicking the effect of super-granular flows) and transport profiles like meridional circulation $v(\theta)$ and differential rotation $\omega(\theta)$. C-PhoTraM employs various finite difference schemes combined with flux limiter algorithms to numerically solve the magnetic induction equation with improved accuracy. In general, the surface flux transport model is the replication of the well-known BL mechanism, which can be expressed by the following mathematical equation:

\begin{eqnarray}
\frac{\partial B_{r}}{\partial t} = - \omega(\theta)\frac{\partial B_{r}}{\partial \phi} - \frac{1}{\mathrm{R_\odot}\,\sin\theta}\frac{\partial }
{\partial \theta}\Big(v(\theta)B_{r}\sin\theta \Big) \nonumber\\ + \frac{\eta}{\mathrm{R_\odot}^{2}}\Bigg[\frac{1}{\sin\theta}\frac{\partial }{\partial \theta}\left(\sin\theta \frac{\partial B_{r}}{\partial \theta} \right)
+ \frac{1}{\sin^{2}\theta}\frac{ \partial^{2} B_{r}}{\partial \phi^{2}}\Bigg] + S(\theta,\phi,t).
\label{eq1}
\end{eqnarray}
In our model, the advective terms, $v(\theta)$ follow an axisymmetric profile, peaking at mid-latitudes with a velocity of 15 m/s and going to zero beyond $\pm75^\circ$, while $\omega(\theta)$ is derived from helioseismic observations, with the profile following \cite{Bhowmik2018, Pal2023}. We set $\eta$ to 250 $\mathrm{Km^2/s}$. Another important term in our model is $S(\theta, \phi, t)$, which is the source term for magnetic flux emergence, accounting for contributions from newly formed sunspot regions based on observational or theoretical input. This term is modeled as an ideal bipolar magnetic region, following the mathematical profile outlined in previous works \citep{Bhowmik2018, Pal2023}.\\
\\
\textbf{Model Source Term:}\\
To generate the source term, we require information on the emergence time, heliographic latitudinal and longitudinal positions, and area of active regions for each solar cycle, all of which are taken from the observational database provided by \cite{rgo}. We record the emergence information for each sunspot when it reaches its maximum area. After 1976, the sunspot records transitioned from the RGO database to the USAF/NOAA database. To ensure consistency between the two datasets, we apply a cross-calibration by multiplying a correction factor of 1.4 to all areas less than 206 microhemispheres for active regions appearing post-1976 \citep{Bhowmik2018}. Other quantities, such as the magnetic flux associated with a sunspot, tilt angle, sunspot group radius, and the separation between the two polarities of a sunspot pair, are derived from the observational data using empirical and mathematical relationships. The magnetic flux ($\Phi$) of each active region is calculated using the relation: $\mathrm{\Phi(A) = 3.5 \times 10^{19} \times A}$, where A is the sunspot area in microhemispheres, and $\Phi$ is the total magnetic flux in Maxwells, whic is assumed to be equally distributed in the two polarities. We estimate the radius of the sunspot from the area information and consider that the radial separation between two polarities of a sunspot pair is proportional to the radius of the sunspot. We compute the standard tilt angle using the square root relation, $\mathrm{\gamma = C\,T_n \sqrt{\lambda}}$ to incorporate Joy's tilt law in a sunspot pair. Here, $\gamma$ is the tilt angle, $\lambda$ is the latitudinal position of the centroid of the AR and $\mathrm{T_n}$ is the tilt coefficient for $\mathrm{n^{th}}$ solar cycle. The variation in tilt angles with solar cycle strength (known as tilt quenching) is modeled through $\mathrm{T_n = 1.73\,-\,0.0035\, S_n}$, as derived from \cite{Jiang2011A&A}, where $\mathrm{S_n}$ denotes the solar cycle strength. The constant factor $\mathrm{C}$ is set to 0.7, accounting for the reduction in tilt angles caused by near-surface localized inflows \citep{Jiang2014}. \\
\\
\textbf{Introducing Stochaticity in the Source Term:}\\
To introduce tilt scatter, we adopt a methodology commonly used in previous studies \citep{Jiang2014, Jiao2021}. These studies show that the tilt angles of the sunspots deviate from Joy’s Law in a manner consistent with a Gaussian distribution. This distribution has a zero mean value, indicating no systematic bias, while its standard deviation ($\sigma$) depends on the sunspot area (A) according to the empirical relation: $\mathrm{\sigma = -11 \times \log(A)+35}$ (For a detailed illustration, see the Appendix section in \cite{Lemerle2015}). For each sunspot group, we randomly sample a tilt scatter value ($\epsilon$) from this Gaussian distribution (having zero mean and $\sigma$ standard deviation) which is determined by that sunspot group's area obtained from the observed database. The resulting tilt angle is then given by $\mathrm{\gamma = C\,T_n \sqrt{\lambda} + \epsilon}$. In our simulations, incorporation of random tilt-angle scatter gives rise to anomalous sunspots, including anti-Joy regions (specially those which carry opposite tilt orientations to standard Joy ARs) and ARs with unusually large tilt angles. We also consider the anti-Hale ARs are distributed randomly all over the cycle phase, emergence latititude and two hemispheres. These anti-Hale sunspot distributions are generated from independent random realizations following the methodology described by \cite{Pal2023}. In Figure \ref{fig:a1}, such anomalous AR distributions (Latitudinal and flux distribution) for solar cycle 18 and 19 are shown.

To quantify the contribution from stochasticity, we calculate the flux-weighted tilt coefficient, $\mathrm{T_A}$ = ($\mathrm{\frac{\sum_i \gamma_i \times \phi_i}{\sum_i \phi_i |\lambda_i|}}$), where $\gamma_i$, $\phi_i$, and $|\lambda_i|$ represent the tilt angle, flux content, and the absolute latitude of the $\mathrm{i^{th}}$ sunspot. When this coefficient is multiplied by the signed total flux emergence during that cycle this generates $\mathrm{\Phi T_A}$ -- which provides physical insight into the average contribution of tilted sunspots to the resulting polar flux at the end of that cycle \citep{Dasi2010, Jiao2021}. \\
\\
\\
\textbf{Estimation of Large-Scale Magnetic Fields:}
\\
We calculate the axial dipole moment and polar flux using the photospheric map generated by the C-PhoTraM simulations utilizing the following relations.

\begin{equation}
\mathrm{DM}(t) = \frac{3}{4\pi R_\odot^{2}}\:\int_{\phi=0}^{2\pi}\:\int_{\lambda = -\pi/2}^{\pi/2} \: \:B_r(\lambda,\phi,t)\:\sin\lambda\:\cos\lambda\:d\lambda\:d\phi,
\end{equation}

\begin{equation}
\Phi_{N/S}(t) = \int_{\phi=0}^{2\pi}\:\int_{\lambda_{N/S}}^{} \: R_\odot^{2}\:B_r(\lambda_{N/S},\phi,t)\:\cos\lambda_{N/S}\:d\lambda_{N/S}\:d\phi
\label{eq6}
\end{equation}
Here $\lambda$ and $\phi$ represent latitude and longitude, respectively.  $R_\odot$ is the solar radius. $\Phi_{N/S}$(t) denotes the polar flux in the northern and southern hemispheres, respectively, and DM(t) is the global axial dipole moment. The polar field is calculated based on the surface magnetic field distribution only in the polar cap region ($\pm$70$^{\circ}$ to $\pm$90$^{\circ}$), whereas the axial dipole moment corresponds to the entire photospheric magnetic field.\\
\\
\textbf{Optimization of the Polar Flux:}
\\
The observational magnetograms and polar field measurements are only available from 1974 onward. To bridge this gap, we use polar flux estimates derived from Mount Wilson Observatory calibrated polar faculae data, which extend back to 1906 \citep{Munoz2012}. We first generate multiple ensembles of AR input source terms for each solar cycle for Monte Carlo (MC) C-PhoTraM simulation. In each ensemble of this
MC simulation, our methodology automatically introduces stochastic variation in the tilt angles and in the distribution of anti-Hale regions in terms of flux content, emergence locations, and phase of appearance. Then we compare the SFT simulated polar flux generated from each ensemble run with observed polar flux values and select a set which satisfies both of the following criteria: (1) the simulated polar flux reversal timing must be within $\pm1$ year of the observed reversal time and (2) the simulated polar flux at the end of each solar cycle must lie within a $\pm20\%$ range of the observed polar flux value. For each sunspot cycle, we identify the optimal solution from the reduced ensembles by selecting the one that best matches the observed polar flux at the cycle minimum. This is how we optimize the simulated polar flux for our study.

\begin{figure}
\centering
\includegraphics[scale=0.42]{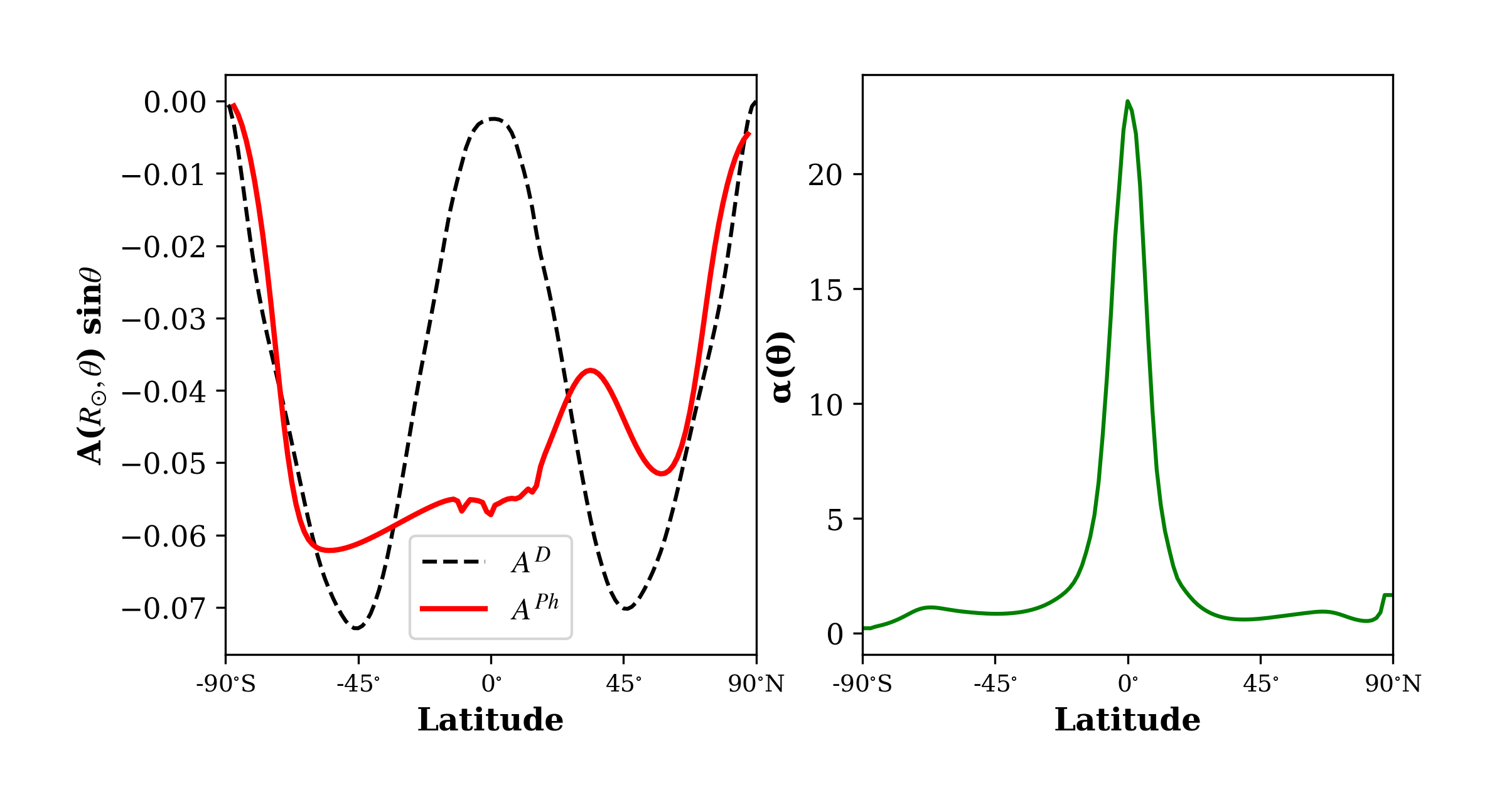}
\caption{The left panel presents surface vector potential at the minima of solar cycle 16 (begining of cycle 17) from C-PhoTraM
simulation ($\mathrm{A^{Ph}}$ denoted by red curve) and from dynamo run ($\mathrm{A^{D}}$ denoted by black dashed line). The right
panel shows the latitudinal correction function $\alpha(\theta)$.}\label{fig:a2}
\end{figure}

\subsection{Dynamo Simulation}\label{sec:methoddynamo}
\textbf{Model Equation:}\\
To generate solar cycles, we utilize the two-dimensional kinematic axisymmetric dynamo model that solves the toroidal (B) and poloidal component ($\mathrm{B_P}$) of the magnetic induction equation in presence of source terms for sustaining the dynamo mechanism \citep{Nandy2002, Chatterjee2004}. Inducing a toroidal field from the poloidal field, followed by the regeneration of the poloidal field from the toroidal field, lies at the heart of the dynamo cycle. In this axisymmetric model, the temporal evolution of the vector potential for the poloidal component $\mathrm{A^D}$(r, $\theta$) and the toroidal component of the magnetic field B(r, $\theta$) are described by the following equations:

\begin{eqnarray}
    \frac{\partial A^D}{\partial t} + \frac{1}{s}\left[ \mathbf{v_p} \cdot \nabla (sA^D) \right] &=& {\eta}_p\left( \nabla^2 - \frac{1}{s^2}  \right)A^D + \alpha B \nonumber\\
    \frac{\partial B}{\partial t}  + s\left[ \mathbf{v_p} \cdot \nabla\left(\frac{B}{s} \right) \right] + (\nabla \cdot \mathbf{v_p})B &=&{\eta}_t\left( \nabla^2 - \frac{1}{s^2}  \right)B \nonumber\\  + s\left(\left[ \nabla \times (A^D \bf \hat{e}_\phi) \right]\cdot \nabla \Omega\right)
    &+& \frac{1}{s}\frac{\partial (sB)}{\partial r}\frac{\partial {\eta}_t}{\partial r}
    \label{eq2}
\end{eqnarray} 

\noindent where, s = r sin$\theta$. In this equation, $\mathrm{v_p}$ represents the poloidal velocity in the meridional plane (or meridional circulation) that causes advection and distortions in the magnetic field. We utilize a single-cell flow in each hemisphere, threading the convection zone. The terms involving $\mathrm{{\eta}_P}$ and $\mathrm{\eta_t}$ correspond to the diffusion of poloidal and toroidal magnetic fields, respectively. 

Source terms in a dynamo model are crucial to compensate for the dissipation of magnetic fields within the convection zone. The internal rotational shear, as determined from helioseismic observations, acts as a source for generating toroidal fields. We adopt the analytical advective and diffusive flow profiles from \cite{Passos2014} for this study. In our simulation, the generation of the poloidal field from the toroidal component is represented by two distinct mechanisms which are thought to operate within the Sun's interior. The first one is the BL Mechanism introduced confined in the near-surface layers. The second mechanism, the mean-field $\alpha$-effect due to helical turbulent convection, operates on relatively weaker toroidal field in the bulk of the solar convection zone. These two terms are incorporated in the equation as a single source term $\alpha$. The mathematical profile with appropriate quenching of these source terms is included following \cite{Passos2014}. In each ensemble of MC simulation, our methodology automatically introduces variation in the tilt angles and in the distribution of anti-Hale regions in terms of flux content, emergence locations, and phase of appearance. \\
\\
\textbf{SFT Maps Coupled with Dynamo Framework:}\\
We follow the methodology described in \cite{Bhowmik2018} to couple the dynamo model with SFT simulations. We emphasize that our model does not incorporate any intrinsic amplitude fluctuation of the $\alpha$ mechanism over time. Instead, the variability in the poloidal field source term is introduced exclusively by integrating the C-PhoTraM generated surface magnetic map at every cycle minimum \citep{Bhowmik2018}. Our C-PhoTraM model generates the radial magnetic field $\mathrm{B^{Ph}_r(R_{\odot},\theta, t_{m})}$ which is related to the magnetic vector potential $\mathrm{A^{Ph}(R_{\odot},\theta,t_{m})}$ through the relation $\mathrm{B = \nabla \times A}$, which reduces to:
\begin{eqnarray}
B^{Ph}_r(R_{\odot},\theta,t_{m})=\dfrac{1}{R_{\odot} \sin \theta} \dfrac{\partial}{\partial \theta} [\sin \theta A^{Ph}(R_{\odot},\theta,t_{m})]
\end{eqnarray}
From this equation, we compute the surface vector potential $\mathrm{A^{Ph}}$ by integrated $\mathrm{B^{Ph}_r(R_{\odot},\theta, t_{m}}$) separately for the two hemispheres. The governing equations are:
\begin{eqnarray}
  A^{Ph}(R_{\odot},\theta,t_{m}) \sin \theta =
  \begin{cases}
    For\,\,\, 0 < \theta < \pi / 2,\\
    \int_{0}^{\theta} B_r(R_{\odot},\theta^{\prime},t_{m})\hspace{0.3cm} \sin \theta^{\prime} d\theta^{\prime} \\ 
    \\ \,
   For\,\,\, \pi/2 < \theta < \pi,\\
   \int_{\theta}^{\pi} B_r(R_{\odot},\theta^{\prime},t_{m})\hspace{0.3cm} \sin \theta^{\prime} d\theta^{\prime} 
  \end{cases}
\end{eqnarray}
It is noted that, here, $\mathrm{B^{Ph}_r(R_{\odot},\theta, t_{m}}$) is calculated by longitudinally averaging the surface magnetic field during solar minima ($\mathrm{t={t_m}}$).

After calculating the vector potential from C-PhoTraM run, denoted as $\mathrm{A^{Ph}}$, it is fed into the dynamo model after proper calibration, which replaces the dynamo-generated vector potential $\mathrm{A^{D}}$ (see left panel of Figure \ref{fig:a2} for an example). To calibrate $\mathrm{A^{D}}$ at solar minima against $\mathrm{A^{Ph}}$, we scale their amplitudes on the solar surface by a constant factor ($k$), which is determined at the minimum of solar cycle 16 and kept fixed throughout the simulation. Thus we use $\mathrm{A^{D}}$ = $k\mathrm{ \times A^{Ph}}$.

After this amplitude calibration, although the amplitudes of $\mathrm{A^{D}}$ and $\mathrm{A^{Ph}}$ are aligned, their latitudinal distributions, specifically $\mathrm{A^{D}(R_{\odot}, \theta, t_{m})\sin\theta}$ and $\mathrm{A^{Ph}(R{\odot}, \theta, t_{m})\sin\theta}$ still differ significantly. To reconcile this, $\mathrm{A^{D}}$ at each solar minimum is further corrected by a latitudinal function $\alpha(\theta)$, such that the product $\mathrm{\alpha(\theta) \times A^{D}(R{\odot}, \theta, t_{m})\sin\theta}$ matches $\mathrm{A^{Ph}(R{\odot}, \theta, t_{m})\sin\theta}$ at the solar surface (see right panel of Figure \ref{fig:a2} for an example). We assume that the correction to the dynamo’s poloidal field -- arising from the BL mechanism -- is confined to the radial range between 0.8R${\odot}$ and R${\odot}$. This assimilation is implemented sequentially at each cycle minima and forces the dynamo generation of the toroidal component of the magnetic field which acts as the source of the sunspot cycle. To elaborate, at each solar minimum, the dynamo simulation is paused, and $\mathrm{A^{D}(r, \theta, t_{m})\sin\theta}$ is multiplied by $k \alpha(\theta)$ at all grid points within this mentioned radial layer (remember $k$ is fixed but $\alpha(\theta)$ changes for each cycle). Once $\mathrm{A^{D}}$ has been updated to reflect the C-PhoTraM-generated vector potential $\mathrm{A^{Ph}}$, the dynamo simulation resumes and continues its evolution. It is to be noted that all the other plasma parameters in the dynamo model are kept fixed for each solar cycle simulation.

This iterative approach enables us to produce a sunspot number time series driven by the C-PhoTraM model, thereby providing a physics-based reconstruction of solar cycles over an extended period, including the extreme variations.

\end{appendix}

\end{document}